\documentclass[reprint,twocolumn,showpacs,amssymb,amsmath,aps,pra]{revtex4-1}
\usepackage{graphics}
\usepackage{bbold}
\usepackage{graphicx}
\usepackage{bm}
\usepackage{amsmath}
\usepackage{amssymb}
\begin{document}

\title{System-time entanglement in a discrete time model}
\author{A.\ Boette, R.\ Rossignoli, N.\ Gigena, M.\ Cerezo}
\affiliation{Instituto de F\'{\i}sica de La Plata and Departamento de F\'{\i}sica,
Universidad Nacional de La Plata, C.C. 67, La Plata (1900), Argentina}

\begin{abstract}
We present a model of discrete quantum evolution based on quantum correlations
between the evolving system and  a reference quantum clock system. A quantum
circuit for the model is provided, which in the case of a constant Hamiltonian
is able to represent the evolution over $2^n$ time steps in terms of just $n$
time qubits and $n$ control gates. We then introduce the concept of system-time
entanglement as a measure of distinguishable quantum evolution, based on the
entanglement between the system and the reference clock. This quantity vanishes
for stationary states and is maximum for systems jumping onto a new orthogonal
state at each time step. In the case of a constant Hamiltonian leading to a
cyclic evolution it is a measure of the spread over distinct energy
eigenstates, and satisfies an entropic energy-time uncertainty relation. The
evolution of mixed states is also examined. Analytical expressions for the
basic case of a qubit clock, as well as for the continuous limit in the
evolution between two states, are provided.
\end{abstract} 
\maketitle

\section{Introduction}
Ever since the foundations of quantum mechanics, time has been mostly
considered as an external classical parameter. Various attempts to incorporate
time in a fully quantum framework have  nonetheless been made, starting with
the Page and Wootters mechanism \cite{PaW.83} and other subsequent proposals
\cite{CR.91,IS.94}. This subject has recently received increasing attention in
both quantum mechanics \cite{M.14,M.15,FC.13,Ve.14,CR.15} and  general
relativity \cite{QT.15,Ga.09}, where this problem is considered a key issue in
the connection between both theories. In the present work we introduce a simple
discrete quantum model of evolution, which on one hand, constitutes a
consistent discrete version of the formalism of \cite{PaW.83,QT.15}, while on
the other hand, provides a practical means to simulate quantum evolutions. We
show that a quantum circuit for the model can be constructed,  which in the
case of  a constant Hamiltonian is able to simulate the evolution over $N=2^n$
times in terms of just $n$ time-qubits and $O(n)$ gates, providing the basis
for a parallel-in-time simulation.

We then introduce and discuss the concept of system-time entanglement, which
arises naturally in the present scenario, as a  quantifier of the actual
distinguishable evolution undergone by the system. Such quantifier can be
related to the minimum 	time necessarily elapsed by the system. For a constant
Hamiltonian we show that  this entanglement is bounded above by the entropy
associated with the spread over energy eigenstates of the initial state,
reaching this bound for a spectrum  leading to a cyclic evolution, in which
case it satisfies an entropic energy-time uncertainty relation. Illustrative
analytical results for a qubit-clock, which constitutes the basic building
block in the present setting, are provided. The continuous limit for the
evolution between two arbitrary states is also analyzed.

\section{Formalism}
\subsection{History states}
We consider a bipartite system $S+T$, where $S$  represents a quantum system
and $T$ a quantum clock system with finite Hilbert space dimension $N$. The
whole system is assumed to be in a pure state of the form
\begin{equation} |\Psi\rangle=\frac{1}{\sqrt{N}}\sum_{t=0}^{N-1}
 |\psi_t\rangle|t\rangle\label{1}\,,\end{equation}
where $\{|t\rangle,\;t=0,\ldots,N-1\}$ is
an orthonormal basis of  $T$ and $\{|\psi_t\rangle,\;t=0,\ldots,N-1\}$
arbitrary pure states of $S$. Such state can describe, for instance, the whole
evolution of an initial pure state $|\psi_0\rangle$ of $S$ at a
discrete set of times $t$. The state $|\psi_t\rangle$ at time $t$ can be
recovered as the conditional state of $S$ after a local measurement at $T$ in
the previous basis with result $t$:
\begin{equation} |\psi_t\rangle\langle\psi_t|=\frac{{\rm
Tr}_T\,[|\Psi\rangle\langle\Psi|\,\Pi_t]}{ \langle\Psi|\Pi_t|\Psi\rangle}\,,\end{equation}
where $\Pi_t=\mathbb{1}\otimes |t\rangle\langle t|$. In shorthand notation
$|\psi_t\rangle\propto \langle t|\Psi\rangle$.

If we write
 \begin{equation} |\psi_t\rangle=U_t|\psi_0\rangle,\;\;t=0,\ldots,N-1\label{u}\,,
 \end{equation}
where $U_t$ are unitary operators at $S$ (with $U_0=\mathbb{1}$), the state
(\ref{1}) can be generated  with the schematic quantum circuit of
Fig.\ \ref{1}. Starting from the product initial state
$|\psi_0\rangle|0\rangle$, a Hadamard-like gate \cite{NC.00} at $T$ turns it
into the superposition
$\frac{1}{\sqrt{N}}\sum_{t=0}^{N-1}|\psi_0\rangle|t\rangle$, after which a
control-like gate $\sum_t U_t\otimes |t\rangle\langle t|$ will transform it  in
the state (\ref{1}). A specific example will be provided in Fig.\ \ref{f2}.

\begin{figure}[htp]
\includegraphics[scale=1]{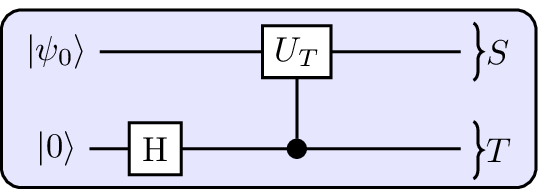}
\caption{(Color online) Schematic circuit representing the generation of the
system-time pure state (\ref{1}). The control gate performs the operation $U_t$
on $S$ if $T$ is in state $|t\rangle$, while the Hadamard-type gate $H$ creates
the superposition $\propto\sum_{t=0}^{N-1}|t\rangle$.} \label{f1}
\end{figure}

From  a formal perspective, the state (\ref{1}) is a ``static'' eigenstate of
the $S+T$ translation ``super-operator''
 \begin{equation} {\cal U}=\sum_{t=1}^{N} U_{t,t-1}\otimes |t\rangle\langle {t-1}|\,,
 \label{H}\end{equation}
where $U_{t,t-1}=U_t U_{t-1}^\dagger$ evolves the state of $S$ from $t-1$ to
$t$ ($|\psi_t\rangle=U_{t,t-1}|\psi_{t-1}\rangle$) and the cyclic condition
$|N\rangle\equiv |0\rangle$, i.e.\ $U_{N,N-1}=U_{N-1}^\dagger$, is imposed.
Then,
\begin{equation} {\cal U}|\Psi\rangle=|\Psi\rangle\label{U1}\,\end{equation}
showing that the state (\ref{1}) remains strictly invariant under such global
translations in the $S+T$ space.

Eq.\ (\ref{U1}) holds for {\it any} choice of initial state
$|\psi_0\rangle$ in (\ref{1}). The eigenvalue $1$ of ${\cal U}$ has then a
degeneracy equal to the Hilbert space dimension $M$ of $S$, since for
$M$ orthogonal initial states $|\psi_0^j\rangle$,
$\langle\psi_0^j|\psi_0^l\rangle=\delta_{jl}$, the ensuing states
$|\Psi^l\rangle$ are orthogonal due to  Eq.\ (\ref{u}):
\begin{equation}\langle\Psi^{l}|\Psi^j\rangle=\frac{1}{N}\sum_{t=0}^{N-1}\langle
\psi^l_{t}|\psi^j_t\rangle=\langle\psi_0^l|\psi_0^j\rangle=\delta_{lj}\,.\end{equation}

The remaining eigenstates of ${\cal U}$ are of the form
$|\Psi_k\rangle=\frac{1}{\sqrt{N}}\sum_{t=0}^{N-1} e^{i2\pi
kt/N}|\psi_t\rangle|t\rangle\label{lk}$  with $k$ integer and represent
the evolution associated with operators  $U_t^k= e^{i2\pi kt/N}U_t$:
 \begin{equation} {\cal U}|\Psi_k\rangle=e^{-i2\pi k/N}|\Psi_k\rangle\,,\;\;k=0,\ldots,N-1
 \label{ck}\,. \end{equation}
All eigenvalues $\lambda_k=e^{-i2\pi k/N}$ are $M$-fold degenerate by the same
previous arguments. The full set of $N$ eigenvalues and a choice of $M N$
orthogonal eigenvectors of ${\cal U}$ are thus obtained. We may then write, for
general $U_t$,
 \begin{equation} {\cal U}=\exp[-i {\cal J}]\label{J}\,,\end{equation}
with ${\cal J}$ hermitian and satisfying ${\cal
J}|\Psi_k\rangle=2\pi\frac{k}{N}|\Psi_k\rangle$ for $k=0,\ldots,N-1$.
In particular, the states (\ref{1}) satisfy
 \begin{equation} {\cal J}|\Psi\rangle=0\,,\end{equation}
which represents a discrete counterpart of the Wheeler-DeWitt equation
\cite{QT.15,DW.67,HH.83} determining the state $|\Psi\rangle$ in continuous
time theories \cite{QT.15}. In the limit where $t$ becomes a continuous
unrestricted variable, the state (\ref{1}) with condition (\ref{u}) becomes in
fact that considered in \cite{QT.15}. Note, however, that here  ${\cal J}$ is
actually defined just modulo $N$, as any ${\cal J}$ satisfying ${\cal
J}|\Psi_k\rangle=2\pi(\frac{k}{N}+n_k)|\Psi_k\rangle$ with $n_k$ integer
will also fulfill Eq.\ (\ref{J}).

All $|\Psi_k\rangle$ are also eigenstates of the hermitian operators ${\cal
U}_{\pm}=i^{\frac{1\mp 1}{2}}({\cal U}\pm{\cal U}^\dagger)/2$, with eigenvalues
$\cos\frac{2\pi k}{N}$ and $\sin\frac{2\pi k}{N}$ respectively, i.e. $1$ and
$0$ for the states (\ref{1}). The latter can then be also obtained as ground
states of $-{\cal U}_+$. An hermitian operator ${\cal H}$ similar to $-{\cal
U}_+$ but with no cyclic condition (${\cal H}=-\tilde{\cal U}_++I_{S}\otimes
I_{T}$, with $\tilde{\cal U}={\cal U}-U^\dagger_{N-1}|0\rangle\langle N-1|
+\frac{1}{2}I_S\otimes(|0\rangle\langle 0|+|N-1\rangle\langle N-1|)$ was
considered in \cite{FC.13} for deriving a variational approximation to the
evolution.

\subsection{Constant evolution operator}
\begin{figure}[htp]
\includegraphics[scale=.7]{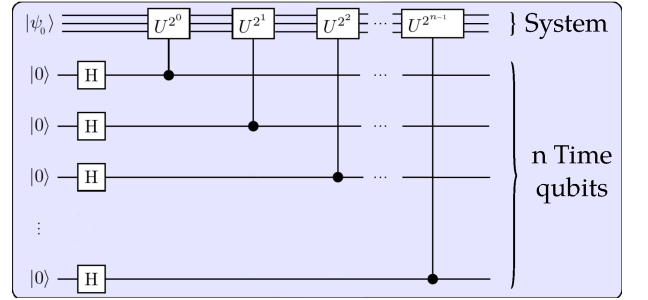}
\caption{(Color online) Circuit representing the generation of the system-time
state (\ref{1}) for $U_t=(U)^t$ and $N=2^n$. The $n$ control gates perform the
operation $U^t=U^{\sum_{j=1}^n t_j 2^{j-1}}$ on the system after writing $t$ in
the binary form $t=\sum_{j=1}^{n} t_j 2^{j-1}$,  while the $n$ Hadamard gates
lead to a coherent sum over all values of the $t_j$'s, i.e., over all $t$ from
$0$ to $2^n-1$.} \label{f2}
\end{figure}

If $U_{t,t-1}=U$ $\forall$ $t$, then
 \begin{equation} U_t=(U)^t=\exp[-iHt]\,,\;\;\;t=0,\ldots,N-1\,,  \label{C}\end{equation}
where $H$ represents a constant Hamiltonian for system $S$. In this case the state
(\ref{1}) can be generated with the first step of the circuit employed for
phase estimation \cite{NC.00}, depicted in Fig.\ \ref{f2}. If $N=2^n$, such
circuit, consisting of just  $n$ time qubits and $m=\log_2 M$  system qubits,
requires only $n$ initial single qubit Hadamard gates on the time-qubits if
initialized  at $|0\rangle$ (such that $|0\rangle_T\equiv
\otimes_{j=1}^n|0_j\rangle\rightarrow
\otimes_{j=1}^n\frac{|0_j\rangle+|1_j\rangle}{\sqrt{2}}=
\frac{1}{\sqrt{N}}\sum_{t=0}^{N-1}|t\rangle$ for $t=\sum_{j=1}^{n}t_j
2^{j-1}$), plus $n$ control $U^{2^{j-1}}$ gates acting on the system qubits,
which perform the operation $U^t|\psi_0\rangle= \prod_{j=1}^n U^{t_j
2^{j-1}}|\psi_0\rangle$. A measurement of the time qubits with result $t$ makes
$S$ collapse to the state $|\psi_t\rangle=e^{-i Ht}|\psi_0\rangle$.

In addition, if $U$ in (\ref{C}) satisfies the cyclic condition
$U^N=\mathbb{1}$, which implies that $H$ should have eigenvalues $2\pi k/N$
with $k$ integer, Eq.\ (\ref{H}) can be written as
 \begin{equation} {\cal U}=U\otimes V=\exp[-i(H\otimes\mathbb{1}_T+\mathbb{1}_S\otimes P)]\,,
 \label{UU}\end{equation}
where $V=\exp[-iP]=\sum_{t=1}^{N} |t\rangle\langle t-1|$ is the  (cyclic) time
translation operator. Its eigenstates are the discrete Fourier transform (FT)
of the time states $|t\rangle$,
\begin{equation} V|\tilde{k}\rangle=e^{-i2\pi k/N}|\tilde{k}\rangle,\;\;
|\tilde{k}\rangle=\frac{1}{\sqrt{N}}\sum_{t=0}^{N-1}e^{i2\pi kt/N}|t\rangle\label{TT}\,,
\end{equation}
for $k=0,\ldots,N-1$, such that $P$ is the ``momentum''
associated with the time operator $T$:
 \begin{equation} T|t\rangle=t|t\rangle\,,\;\;P|\tilde{k}\rangle=2\pi
 \frac{k}{N}|\tilde{k}\rangle\,.\end{equation}
Hence, ${\cal J}=H\otimes\mathbb{1}_T+\mathbb{1}_S\otimes P$ adopts in this
case the same form as that of continuous theories \cite{QT.15}.

\subsection{System-Time entanglement}
Suppose now that one wishes to quantify consistently the ``amount''  of
distinguishable evolution of a pure quantum state. Such measure can be related
to a minimum time $\tau_m$ (number or fraction of steps) necessarily elapsed by
the system. If the state is stationary, $|\psi_t\rangle\propto |\psi_0\rangle$
$\forall$ $t$, the quantifier should vanish (and $\tau_m=0$) whereas if all $N$
states $|\psi_t\rangle$ are orthogonal to each other, the quantifier should be
maximum (with $\tau=N-1$), indicating that the state has indeed evolved through
$N$ distinguishable states. We now propose the entanglement of the pure state
(\ref{1}) (system-time entanglement) as such quantifier, with $\tau_m$ an
increasing function of this entanglement. In Figs.\ \ref{f1}--\ref{f2}, such
entanglement is just that between the system and the time-qubits, generated by
the control $U_t$. 

We first note that Eq.\ (\ref{1}) is not, in general, the Schmidt decomposition
\cite{NC.00} of the state $|\Psi\rangle$, which is
 \begin{equation} |\Psi\rangle=\sum_{k}\sqrt{p_k}|k\rangle_S|k\rangle_T\label{sd}\,,\end{equation}
where $|k\rangle_{S(T)}$ are orthogonal states of $S$ and $T$ ($_\mu\langle
k|k'\rangle_{\mu}=\delta_{kk'}$) and $p_k$  the eigenvalues
of the reduced states of $S$ and $T$,
\begin{equation} \rho_{S(T)}={\rm Tr}_{T(S)}|\Psi\rangle\langle\Psi|=
\sum_k p_k|k\rangle_{S(T)}\langle k|\,.
\end{equation}
The entanglement entropy between $S$ and $T$ is then
\begin{equation}
E(S,T)=S(\rho_S)=S(\rho_T)=-\sum_k p_k\log_2 p_k\,,\label{2}
\end{equation}
where $S(\rho)=-{\rm Tr}\rho\log_2\rho$ is the von Neumann entropy.

Eq.\ (\ref{2}) satisfies the basic requirements of an evolution quantifier. If
the state of $S$ is stationary, $|\psi_t\rangle=e^{i\gamma_t}|\psi_0\rangle$ $\forall$
$t$, the state (\ref{1}) becomes {separable},
\begin{equation} |\Psi\rangle=|\psi_0\rangle(\frac{1}{\sqrt{N}}\sum_t
 e^{i\gamma_t}|t\rangle)\,,\label{sta}\end{equation}
implying $E(S,T)=0$. In contrast, if $|\psi_t\rangle$ evolves through $N$
orthogonal states, then $|\Psi\rangle$ is {\it maximally entangled}, with Eq.\
(\ref{1}) already its Schmidt decomposition  and
\begin{equation} E(S,T)=E_{\rm max}(S,T)=\log_2 N\label{Emax}\,.\end{equation}
It is then natural to define the minimum  time $\tau_m$ as
\begin{equation}\tau_m=2^{E(S,T)}-1\,,\end{equation}
which takes the values $0$ and $N-1$ for the previous extreme cases.
The vast majority of evolutions will lie in
between. For instance, a periodic evolution of period $L<N$ with $N/L$ integer,
such that $|\psi_{t+L}\rangle=e^{i\gamma}|\psi_{t}\rangle$ $\forall$ $t$,  will lead to
\begin{equation} |\Psi\rangle=\frac{1}{\sqrt{L}}\sum_{t=0}^{L-1}
|\psi_t\rangle|t_L\rangle,\;\;
|t_L\rangle=\sqrt{\frac{L}{N}}\sum_{k=0}^{N/L-1}e^{i\gamma k}|t+L k\rangle
 \label{L}\,,\end{equation}
with $\langle t'_L|t_L\rangle=\delta_{tt'}$. Hence, its entanglement
$E(S,T)$ {\it will be the same as that obtained with an  $L$ dimensional
effective clock}, as it should. Its maximum value, obtained for $L$
orthogonal states, will then be $\log_2 L$, in which case $\tau_m=L-1$.

The Schmidt decomposition (\ref{sd}) represents in this context the ``actual''
evolution between orthogonal states, with $p_k$ proportional to the
``permanence time'' in each of them. A measurement on $T$ in the Schmidt Basis
would always identify orthogonal states of $S$ for different results (and
viceversa), with the  probability distribution of results indicating the
``permanence'' in these states. If in Eq.\ (\ref{1}) there are $n_k$ times $t$
where $|\psi_t\rangle\propto |k\rangle_S$,  with $\sum_k n_k=N$ and
$|k\rangle_S$ orthogonal states, then  
\[|\Psi\rangle=\sum_k \sqrt{\frac{n_k}{N}}|k\rangle_S
(\frac{1}{\sqrt{n_k}}\sum_{t/|\psi_t\rangle\propto|k\rangle_S}e^{i\gamma_t}|t\rangle)\,,\]
which is the Schmidt decomposition (\ref{sd}) with $p_k\propto n_k$,
i.e. proportional to the total time in the state $|k\rangle_S$.
Note also that Eqs.\ (\ref{sd})--(\ref{2})  are essentially symmetric, so that the roles of
$S$ and $T$ can in principle be interchanged.

{\it Quadratic entanglement.} A simple quantifier for the general case can be
obtained through the entanglement determined by the  entropy
$S_2(\rho)=2(1-{\rm Tr}\,\rho^2)$, which is just a linear function of the
purity ${\rm Tr}\,\rho^2$ and does not require the evaluation of the
eigenvalues of $\rho$ \cite{S2.00,RC.03,GR.14} (purity is also more easily
accessible experimentally \cite{T.13}).  We obtain, using
$\rho_S=\frac{1}{N}\sum_t |\psi_t\rangle\langle\psi_t|$,
\begin{eqnarray}E_2(S,T)&=&S_2(\rho_T)=S_2(\rho_S)=2(1-{\rm Tr}\,\rho_S^2)\nonumber\\
&=&2{\textstyle\frac{N-1}{N}}(1-{\textstyle\frac{1}{N(N-1)}}\sum_{t\neq
 t'}|\langle\psi_t|\psi_{t'}\rangle|^2)\,,\label{E2}\end{eqnarray}
which is just a decreasing function of the average pairwise squared fidelity
between all visited  states. If they are all proportional, $E_2(S,T)=0$ whereas
if they are all orthogonal, $E_2(S,T)=2\frac{N-1}{N}$ is maximum. If $S$ and $T$ are qubits
$E_2(S,T)$ is just the squared {\it concurrence} \cite{Wo.97} of
$|\Psi\rangle$.

\subsection{Relation with energy spread}
In the constant case (\ref{C}), we may expand $|\psi_0\rangle$ in the
eigenstates of $U$ or  $H$, $|\psi_0\rangle=\sum_k c_k |k\rangle$ with
$H|k\rangle=E_k|k\rangle $, such that $|\psi_t\rangle=\sum_k c_k e^{-i E_k
t}|k\rangle$  and
 \begin{equation} |\Psi\rangle=\frac{1}{\sqrt{N}}\sum_{k,t} c_k e^{-iE_k t}|k\rangle|t\rangle
 =\sum_k c_k |k\rangle|\tilde{k}\rangle_T\,,\label{CK}\end{equation}
with $|\tilde{k}\rangle_T=\frac{1}{\sqrt{N}}\sum_{t}e^{-iE_k t}|t\rangle$. We
can always assume all $E_k$ distinct in (\ref{CK}) such that $c_k |k\rangle$ is
the projection of $|\psi_0\rangle$ onto the eigenspace with energy $E_k$. In
the cyclic case $U^N=\mathbb{1}$, with $E_k=2\pi k/N$, $k=0,\ldots,N-1$, the
states $|\tilde{k}\rangle_T$ become the orthogonal FT states (\ref{TT}) 	
($|\tilde{k}\rangle_T=|-\tilde{k}\rangle$). Eq.\
(\ref{CK}) is then {\it the Schmidt decomposition} (\ref{sd}), with
$p_k=|c_k|^2$ and
\begin{equation} E(S,T)=-\sum_k |c_k|^2\log_2 |c_k|^2\label{ESTK}\,.\end{equation}
For this spectrum, entanglement becomes then {\it a measure of the spread of
the initial state $|\psi_0\rangle$ over the eigenstates of $H$} with distinct
energies. The same holds in the quadratic case (\ref{E2}) where
$E_2(E,T)=2\sum_k |c_k|^2(1-|c_k|^2)$. If there is no dispersion
$|\psi_0\rangle$ is stationary and entanglement vanishes while if
$|\psi_0\rangle$ is uniformly spread over $N$ eigenstates it is maximum
($E(S,T)=\log_2 N$).

While Eq.\ (\ref{ESTK}) also holds for a displaced spectrum
$E_k=E_0+2\pi k/N$, for an arbitrary spectrum $\{E_k\}$ it will hold
approximately if the overlaps $_T\langle
\tilde{k}|\tilde{k'}\rangle_T=\frac{1}{N}\sum_t e^{-i(E_k-E_{k'})t}$ are
sufficiently small for $k\neq k'$. In general  we actually have the strict
bound
\begin{equation}E(S,T)\leq -\sum_k |c_k|^2\log_2 |c_k|^2, \label{ESTK2}\end{equation}
since $|c_k|^2=\sum_{k'}p_{k'}|\langle k|k'\rangle_S|^2$, with $|k\rangle$ the
eigenstates of $H$ and $|k'\rangle_S$ the Schmidt states in (\ref{sd}), which
implies that the $|c_k|^2$'s are {\it majorized} \cite{Bha.97} by the $p_k$'s:
\begin{equation}
\{|c_k|^2\}\prec\{p_k\}\,,  \label{prec}\end{equation} where $\{|c_k|^2\}$ and
$\{p_k\}$ denote the sets sorted in decreasing order. Eq.\ (\ref{prec})
(meaning $\sum_{k=1}^j |c_k|^2\leq \sum_{k=1}^jp_k$ for $j=1\,\ldots,N-1$)
implies that the inequality (\ref{ESTK2}) actually holds for any Schur-concave
function of the probabilities \cite{Bha.97}, in particular for any entropic
form  $S_f(\rho)={\rm Tr}\,f(\rho)$ with $f(p)$ concave and satisfying
$f(0)=f(1)=0$ \cite{GR.14,CR.02}, such as the von Neumann entropy
($f(\rho)=-\rho\log_2\rho$) and the previous $S_2$ entropy
($f(\rho)=2\rho(\mathbb{1}-\rho)$):
\begin{equation}E_f(S,T)=\sum_k f(p_k)\leq \sum_k f(|c_k|^2), \label{ESTK3}\end{equation}
as can be easily verified. Eqs.\ (\ref{ESTK})--(\ref{ESTK3}) then indicate that
the entropy of the spread over Hamiltonian eigenstates of the initial state
provides an upper bound to the corresponding system-time entanglement entropy
than can be generated by {\it any} Hamiltonian diagonal in the states
$|k\rangle$. The bound is always reached for an equally spaced spectrum
$E_k=2\pi k/N\in[0,2\pi]$ leading to a cyclic evolution, which therefore
generates {\it the highest possible system-time entanglement for a given
initial spread $\{|c_k|^2\}$}. 

\subsection{Energy-time uncertainty relations}
For the aforementioned equally spaced spectrum, 
we may also expand the state $|\psi_0\rangle$ of $S$ in an
orthogonal set of uniformly spread states,
\begin{equation} {\textstyle|\psi_0\rangle=\sum_{l=0}^N \tilde{c}_l|\tilde{l}\rangle_S\,,
 \;\;|\tilde{l}\rangle_S=\frac{1}{\sqrt{N}}\sum_k e^{i2\pi kl/N}|k\rangle\,,}\end{equation}
with $\tilde{c}_l=\frac{1}{\sqrt{N}}\sum_k e^{-i2\pi k/N}c_k$ the FT of the
$c_k$'s in (\ref{CK}). Since
$U^t|\tilde{l}\rangle_S=|\widetilde{l-t}\rangle_S$, it is verified that these
maximally spread states $|\tilde{l}\rangle_S$ (which according to Eq.\ (\ref{ESTK}) 
lead to maximum system-time entanglement $E(S,T)=\log_2 N$) indeed evolve through $N$
orthogonal states $|\widetilde{l-t}\rangle_S$. Moreover, Eq.\ (\ref{CK}) becomes
\begin{equation} |\Psi\rangle=\sum_{l,t}
\tilde{c}_l|\widetilde{l-t}\rangle_S|t\rangle=\sum_{l}|\tilde{l}\rangle_S(\sum_t
 \tilde{c}_t |t-l\rangle)\,, \label{U}\end{equation}
showing that $\tilde{c}_l$ determines the distribution of time states
$|t\rangle$ assigned to each state $|\tilde{l}\rangle_S$, i.e., the uncertainty
in its time location. Being related through a finite FT, $\{c_k\}$ and
$\{\tilde{c}_l\}$ satisfy various uncertainty relations, such as {\small
\cite{DCT.91,PDO.01,Hi.57}}
\begin{equation} E(S,T)+\tilde{E}(S,T)\geq \log_2 N\label{US}\,,\end{equation}
where $\tilde{E}(S,T)=-\sum_l |\tilde{c}_l|^2\log_2 |\tilde{c}_l|^2$ is the
entropy characterizing the time uncertainty and $E(S,T)$ the energy uncertainty
(\ref{ESTK}). If localized in energy ($|c_{k}|=\delta_{kk'}$, $E(S,T)=0$), Eq.\
(\ref{US}) implies maximum time uncertainty
($|\tilde{c}_l|=\frac{1}{\sqrt{N}}$, $\tilde{E}(S,T)=\log_2 N$) and viceversa.
We also have  $n(\{c_k\})\,n(\{\tilde{c}_l\})\geq N$ \cite{DS.89}, where
$n(\{\alpha_k\})$ denotes the number of non-zero $\alpha_k$'s. Bounds for the
product of variances in the discrete FT are discussed in \cite{MS.08}.

\subsection{Mixed states}
Let us now consider that  $S$ is a bipartite system $A+B$.
By taking the partial trace of  (\ref{1}),
 \begin{equation} \rho_{BT}={\rm Tr}_A\,|\Psi\rangle\langle\Psi|=
 \sum_j {_A}\langle j|\Psi\rangle\langle \Psi|j\rangle_A\label{BT}\,,\end{equation}
we see that the system-time state for a subsystem is a {\it mixed state}. Of
course, the state of $B$ at time $t$, setting now $\Pi_t=I_B\otimes
|t\rangle\langle t|$,  is given by the standard expression
\begin{equation}\rho_{Bt}=\frac{{\rm Tr}_T\, \rho_{BT}\Pi_t}{{\rm Tr}\,\rho_{BT}\Pi_t}
 = {\rm Tr}_A|\psi_t\rangle\langle\psi_t|\label{Bt}\,.\end{equation}
If the initial state of $S$ is
$|\psi_0\rangle=\sum_j\sqrt{q_j}|j\rangle_A|j\rangle_B$ (Schmidt
decomposition), Eqs.\ (\ref{BT})--(\ref{Bt}) determine the evolution of an
initial mixed state $\rho_{B0}=\sum_j q_j |j\rangle_B\langle j|$ of $B$,
considered as a subsystem in a purified state undergoing unitary evolution. For
instance, if just subsystem $B$ evolves, such that $U_t=I_A\otimes U_{Bt}$
$\forall$ $t$, Eq.\ (\ref{BT}) leads to
\begin{eqnarray}\rho_{BT}&=&\sum_j q_j|\Psi_j\rangle_{BT}\langle\Psi_j|\,,\label{BTs}
 \end{eqnarray}
where
$|\Psi_j\rangle_{BT}=\frac{1}{\sqrt{N}}\sum_{t=0}^{n-1}U_{Bt}|j\rangle_B|t\rangle$.
Eq.\ (\ref{Bt}) is then the mixture of the pure $B+T$ states associated with
each eigenstate of $\rho_{B0}$, and implies  the unitary evolution
$\rho_{Bt}=U_{Bt}\rho_{B0}U^\dagger_{Bt}$.

Since the state (\ref{BT}) is in general mixed, the correlations between $T$
and a subsystem $B$ can be more complex than those with the whole system $S$.
The state (\ref{BT}) can in principle exhibit distinct types of correlations,
including entanglement \cite{RF.89,BV.96},  discord-like correlations
\cite{DZ.01,HV.03,KM.12,RCC.10} and  classical-type correlations. The exact
evaluation of the quantum correlations is also more difficult, being in general
a hard problem  \cite{DP.04,YH.14}. We will here consider just the entanglement
of formation \cite{BV.96} $E(B,T)$ of the state (\ref{BT}), which, if nonzero,
indicates that (\ref{BT}) cannot be written as a convex mixture of pure product
states \cite{RF.89}
$|\Psi_\alpha\rangle_{BT}=|\psi_\alpha\rangle_{B}|\phi_\alpha\rangle_T$. In
this context the latter represent essentially {\it stationary} states.
Separability with time would then indicate that $\rho_{BT}$ can be written as a
convex mixture of such states, requiring no quantum interaction with the clock
system for its formation.

\section{Examples}
\subsection{The qubit clock}
As  illustration, we examine the basic case of a qubit
clock ($N=2$). Eq.\ (\ref{1}) becomes
\begin{eqnarray}
|\Psi\rangle&=&(|\psi_0\rangle|0\rangle+|\psi_1\rangle|1\rangle)/\sqrt{2}\nonumber\\
&=&\sqrt{p_+}|++\rangle+\sqrt{p_-}|--\rangle\,,\label{sd2}\\
p_{\pm}&=&(1\pm|\langle\psi_0|\psi_1\rangle|)/2\,,\nonumber
\end{eqnarray}
where $|\psi_1\rangle=U|\psi_0\rangle$ and (\ref{sd2}) is its Schmidt
decomposition, with $|\pm\rangle_S=(|\psi_0\rangle\pm e^{-i\gamma}
|\psi_1\rangle)/\sqrt{4p_{\pm}}$, $|\pm\rangle_T=(|0\rangle\pm
e^{i\gamma}|1\rangle)/\sqrt{2}$ and $e^{i\gamma}=\frac{\langle
\psi_0|\psi_1\rangle}{|\langle\psi_0|\psi_1\rangle|}$. Hence,
$E(S,T)=-\sum_{\nu=\pm} p_\nu\log p_\nu$  will be fully determined by the
overlap or {\it fidelity}  $|\langle\psi_0|\psi_1\rangle|$ between the initial
and final states, decreasing as the fidelity increases and becoming maximum for
orthogonal states. The quadratic entanglement entropy $E_2(S,T)$
 becomes just
\begin{equation} E_2(S,T)=4p_+p_-=1-|\langle\psi_0|\psi_1\rangle|^2\label{SF}\,.\end{equation}
These results hold for arbitrary dimension $M$ of $S$.

The operator (\ref{H}) becomes ${\cal U}=U\otimes|1\rangle\langle
0|+U^\dagger\otimes|0\rangle\langle 1|$, and is directly hermitian,  with
eigenvalues $e^{i2k\pi/2}=\pm 1$ for $k=0$ or $1$, $M$-fold degenerate. Hence,
in this case
\begin{equation} {\cal J}=\pi({\cal U}-\mathbb{1})/2\,,\end{equation}
involving coupling between $S$ and $T$ unless $U^\dagger\propto U$.

For $|\psi_1\rangle$ close to $|\psi_0\rangle$, Eq.\ (\ref{SF}) becomes
proportional to the Fubini-Study metric \cite{AA.90}. If $U=\exp[-i\epsilon
h]$, an expansion of $|\psi_0\rangle$ in the eigenstates of $h$,
$|\psi_0\rangle=\sum_k c_k |k\rangle$ with $h|k\rangle=\varepsilon_k
|k\rangle$, leads to
\begin{equation} E_2(S,T)=1-|\sum_k |c_k|^2
e^{-i\epsilon \varepsilon_k}|^2\approx \epsilon^2(\langle h^2\rangle-\langle h\rangle^2)\label{apr}\,,
\end{equation}
where the last expression holds up to $O(\epsilon^2)$. Hence, for a ``small''
evolution the system-time entanglement of a single step is determined by the
energy fluctuation $\langle h^2\rangle-\langle h\rangle^2$ in $|\psi_0\rangle$
($\langle O\rangle\equiv\langle \psi_0|O|\psi_0\rangle$), with $E_2(S,T)$
directly proportional to it. For instance, if $S$ is also a single qubit and
$\varepsilon_1-\varepsilon_0=\varepsilon$, the exact expression becomes
\begin{eqnarray} E_2(S,T)&=&4\sin^2(\frac{\epsilon \varepsilon}{2})|c_0|^2|c_1|^2\label{apr1}\\&=&
4\sin^2(\frac{\epsilon\varepsilon}{2})\frac{\langle h^2 \rangle-\langle
 h\rangle^2}{\varepsilon^2}\,, \label{apr2}\end{eqnarray}
which reduces to (\ref{apr}) for small $\epsilon$. It is also verified that
$E_2(S,T)\leq S_2(|c_0|^2,|c_1|^2)=4|c_0|^2|c_1|^2$, i.e., it is upper bounded
by the quadratic entropy of the energy spread (Eq.\ \ref{ESTK3}), reaching the
bound for $E=\epsilon\varepsilon=\pi$, in agreement with the general result (\ref{ESTK})--(\ref{ESTK2}. 
Returning to the case of a general $S$, we also note that $E_2(S,T)$ determines
the minimum time required for the evolution from $|\psi_0\rangle$ to
$|\psi_1\rangle$ in standard continuous time theories \cite{AA.90}, which
depends on the fidelity $|\langle \psi_0|\psi_1\rangle|$ and can  then be
expressed in terms of $E_2$ as $\hbar\sin^{-1}(\sqrt{E_2(S,T)})/\sqrt{\langle
h^2\rangle-\langle h\rangle^2}$.

Let us now assume that $S=A+B$ is a two qubit-system, with  $U=I_A\otimes U_B$.
As previously stated, starting from an initial entangled pure state of $A+B$
(purification of $\rho_{B0}$), the state (\ref{sd2}) will determine the
evolution of the reduced state of $B$, leading to
\begin{equation}\rho_{Bt}=p|\psi_t^0\rangle\langle\psi_t^0|+q|\psi_t^1\rangle\langle\psi_t^1|\,,
\;\;t=0,1\label{rhob00}\end{equation}
 where $p+q=1$, $\langle\psi_0^0|\psi_0^1\rangle=0$ and $|\psi_1^j\rangle=U_B|\psi_0^j\rangle$ for $j=0,1$.
The reduced state  (\ref{BTs}) of $B+T$ becomes
 \begin{equation} \rho_{BT}=p|\Psi_0\rangle\langle\Psi_0|+q|\Psi_1\rangle\langle\Psi_1|\,,
 \label{rt}\end{equation}
with
$|\Psi_j\rangle=\frac{1}{\sqrt{2}}(|\psi_0^j\rangle|0\rangle+|\psi_1^j\rangle|1\rangle)$.
Since  (\ref{rt}) is a two-qubit mixed state, its entanglement  of formation
can be obtained through the concurrence \cite{Wo.97} $C(B,T)$, whose square is
just the entanglement monotone associated with the quadratic entanglement
entropy $E_2$ ($C^2(B,T)=E_2(B,T)$ for a pure $B+T$ state). It adopts here the
simple expression
 \begin{equation}
C^2(B,T)=(p-q)^2(1-|\langle
\psi_0^j|\psi_1^j\rangle|^2)\label{CBT}\,,\end{equation} where $|\langle \psi_
0^j|\psi_1^j\rangle|=|\langle \psi_0^j|U_B|\psi_0^j\rangle|$ is the same for
$j=0$ or $1$ in a qubit system if $\langle\psi_0^0|\psi_0^1\rangle=0$.   Eq.\
(\ref{CBT}) is then the pure state result (\ref{SF}) for any of the eigenstates
of $\rho_{B0}$ diminished by the factor $(p-q)^2$, vanishing if $\rho_{B0}$ is
maximally mixed ($p=q$). Remarkably, Eq.\ (\ref{CBT}) can be also written as
 \begin{equation} C^2(B,T)=1-F^2(\rho_{B0},\rho_{B1})\,,\label{CF}\end{equation}
where $F(\rho_{B0},\rho_{B1})={\rm
Tr}\,\sqrt{\rho_{B0}^{1/2}\rho_{B1}\rho_{B0}^{1/2}}$  is again the {\it
fidelity} between the initial and final reduced mixed states of $B$
($F=|\langle \psi_0|\psi_1\rangle|$ if $\rho_{B0}$, $\rho_{B1}$ are pure
states). Note also that the total quadratic entanglement entropy is here
\[E_2(S,T)=1-|p\langle \psi_1^0|\psi_0^0\rangle+q\langle \psi_1^1|\psi_0^1\rangle|^2,\]
satisfying $E_2(S,T)\geq C^2(B,T)$ in agreement with the monogamy inequalities
\cite{S2.00,OV.06}, coinciding iff $pq=0$ (pure case).

\subsection{The continuous limit}
Let us now assume that system $S$ is a qubit, with  $T$ of dimension $N$
($t=0,\ldots,N-1$). This case can also represent the evolution  from an initial
state $|\psi_0\rangle$ to an arbitrary final state $|\psi_f\rangle$  in a
general system $S$ of  Hilbert space dimension $M$ if all intermediate states
$|\psi_t\rangle$ belong to the subspace generated by $|\psi_0\rangle$ and
$|\psi_f\rangle$, such that the whole evolution is contained in a
two-dimensional subspace of $S$. Writing the system states as
\begin{equation}|\psi_t\rangle=\alpha_t|0\rangle+\beta_t|1\rangle\,,\;\;t=0,\ldots,N-1,\end{equation}
with $\langle 0|1\rangle=0$ and $|\alpha_t|^2+|\beta_t|^2=1$, we may rewrite state (\ref{1}) as
\begin{eqnarray}|\Psi\rangle&=&\frac{1}{\sqrt{N}}[|0\rangle(\sum_{t}\alpha_t|t\rangle)+
|1\rangle(\sum_{t} \beta_t|t\rangle)]\nonumber\\
&=&\alpha|0\rangle|\phi_0\rangle+\beta|1\rangle|\phi_1\rangle\,,\label{0T}\end{eqnarray}
where 
$|\phi_0\rangle=\frac{1}{\sqrt{N}\alpha}\sum_t\alpha_t|t\rangle$,
$|\phi_1\rangle=\frac{1}{\sqrt{N}\beta}\sum_t\beta_t|t\rangle$, are
normalized (but not necessarily orthogonal) states of $T$ and
all sums over $t$ are from $0$ to $N-1$,  with
\begin{equation}\alpha^2=\frac{1}{N}\sum_{t}|\alpha_t|^2,\;\;\beta^2=\frac{1}{N}\sum_{t}
|\beta_t|^2=1-\alpha^2\,.\label{al}\end{equation}
The Schmidt coefficients of the state (\ref{0T}) are given by
\begin{equation}p_{\pm}=\frac{1}{2}(1\pm\sqrt{1-4\alpha^2\beta^2(1-|\langle\phi_1|\phi_0\rangle|^2)})\,
.\label{ppm2}\end{equation}
 We then obtain
\begin{eqnarray}E_2(S,T)&=&4p_+p_-=4\alpha^2\beta^2(1-|\langle\phi_1|\phi_0\rangle|^2)\nonumber\\
&=&4(\alpha^2\beta^2-\gamma^2),\ \ \gamma=\frac{1}{N}|\sum_{t}\beta_t^*\alpha_t|\label{E22}\,,
\end{eqnarray}
a result which also follows directly from Eq.\ (\ref{E2}).

Let us consider, for instance, the states
\begin{equation}
|\psi_t\rangle=\cos(\frac{\phi t}{N-1})|0\rangle+\sin(\frac{\phi t}{N-1})|1\rangle\,,\end{equation}
such that $S$ evolves from $|\psi_0\rangle=|0\rangle$ to
\[|\psi_f\rangle=\cos\phi|0\rangle+\sin\phi|1\rangle\,,\]
in $N-1$ steps through intermediate equally spaced states contained within the
same plane in the Bloch sphere of $S$.  The $S-T$ entanglement of this $N$-time
evolution can be evaluated exactly with Eqs.\ (\ref{al})--(\ref{E22}), which
yield
\begin{equation}
E_2(S,T_N)=1-\frac{\sin^2\left(\frac{N\phi}{N-1}\right)}{N^2\sin^2\left(\frac{\phi}{N-1}\right)}
\,.\label{E2m}
\end{equation}
For $N=2$ (single step) we recover Eq.\ (\ref{SF})
($E_2(S,T_2)=1-\cos^2\phi=1-|\langle\psi_0|\psi_f\rangle|^2$). If 
$\phi\in[0,\pi/2]$, $E_2(S,T_N)$ is a {\it decreasing} function of $N$ (and an
increasing function of $\phi$),  but rapidly  saturates, approaching {\it a
finite limit for} $N\rightarrow\infty$, namely,
\begin{equation}
E_2(S,T_\infty)=1-\frac{\sin^2\phi}{\phi^2} \,.\label{E3m}\end{equation}
Therefore, system-time entanglement decreases as the number of steps through
intermediate states between $|\psi_0\rangle$ and $|\psi_f\rangle$ is increased,
reflecting the lower average distinguishability between the evolved states, but
remains {\it finite} for $N\rightarrow\infty$. In this limit it is still  an
increasing function of $\phi$ for $\phi\in[0,\pi/2]$, reaching
$1-4/\pi^2\approx 0.59$ for $\phi=\pi/2$, i.e., when the system evolves to an
orthogonal state ($|\psi_f\rangle=|1\rangle$), and reducing to $\approx
\phi^2/3$  for  $\phi\rightarrow 0$. Hence, as compared with a single step
evolution ($N=2$),  the ratio $E_2(S,T_\infty)/E_2(S,T_2)$ increases from $1/3$
for $\phi\rightarrow 0$ to $\approx 0.59$ for $\phi\rightarrow \pi/2$.

If $\phi$ is increased beyond $\pi/2$, the coefficients $\alpha_t$, $\beta_t$
cease to be all positive and entanglement can increase beyond $\approx 0.59$
due to the decreased overlap $\gamma$, reflecting higher average distinguishability 
between evolved states. Entanglement $E_2(S,T_{\infty})$ reaches
in fact $1$ at $\phi=\pi$ (and also $k\pi$, $k\geq 1$ integer), i.e., when the
final state is proportional to the initial state after having covered the whole
circle in the Bloch sphere, since for these values the time states
$|\phi_0\rangle$ and $|\phi_1\rangle$ become orthogonal and with equal weights.
Note also that for $\phi>\pi/2$, $E_2(S,T_N)$ is not necessarily a decreasing
function of $N$, nor an increasing function of $\phi$, exhibiting oscillations:
$E_2(S,T_N)=1$ for $\phi=k\pi(N-1)/N$, $k\neq lN$, and  $E_2(S,T_N)\rightarrow
0$ for $\phi\rightarrow l\pi(N-1)$, $l$ integer.

\section{Conclusions}
We proposed a parallel-in-time discrete model of quantum evolution based on a
finite dimensional clock entangled with the system. The ensuing history state
satisfies a discrete Wheeler-DeWitt-like equation and can be generated through
a simple circuit, which for a constant evolution operator can be efficiently
implemented with just $O(n)$ qubits and control gates for $2^n$ time intervals.

We  then showed that the system-clock entanglement $E(S,T)$ is a measure of the
actual distinguishable evolution undergone by one of the systems relative to
the other.  A natural interpretation of the Schmidt decomposition in terms of
permanence in distinguishable evolved states is also obtained. For a constant
Hamiltonian leading to a cyclic evolution, this entanglement is a measure of
the energy spread of the initial state and satisfies an entropic uncertainty
inequality with a conjugated entropy which measures the time spread. Such
Hamiltonian was rigorously shown to provide the {\it maximum} entanglement
$E(S,T)$ compatible with a given distribution over Hamiltonian eigenstates. For
other Hamiltonians, $E(S,T)$ (and  also general entanglement entropies
$E_f(S,T)$) are strictly bounded by the corresponding entropy of this
distribution.  We have also considered the evolution of mixed states. Although
in this case the evaluation and interpretation of system-clock entanglement and
correlations become more involved, in the simple yet fundamental case of a
qubit clock coupled with a qubit subsystem, such entanglement was seen to be
directly determined by the {\it fidelity} between the initial and final states
of the qubit. A direct relation between this entanglement and energy
fluctuation was also derived for the pure case. Finally, we have also shown
that $E(S,T)$ does remain finite and non-zero in the continuous limit, i.e., 
when the system evolves from an initial to a final state through an arbitrarily 
large number of closely lying equally spaced intermediate states.

The present work opens the way to various further developments, starting from
the definition of proper time basis according to the Schmidt decomposition. It
could be also possible in principle  to incorporate other effects such as
interaction between clocks \cite{CR.15}, explore possibilities of an emergent
space-time or a qubit model for quantum time crystals \cite{Wi.12}. At the very
least, it provides a change of perspective, allowing to identify a qubit clock
as a fundamental ``building block'' of a discrete-time based  quantum
evolution.

\acknowledgments
The authors acknowledge support from CIC (RR) and CONICET (AB,NG,MC) of Argentina.


\begin{thebibliography}{999}
\bibitem{PaW.83} D.N.\ Page, W.K.\  Wootters,  Phys.\ Rev.\ D 27, 2885 (1983);
W.\  Wootters,  Int.\ J.\ Theor.\ Phys.\ 23, 701 (1984).
\bibitem{CR.91} C.\ Rovelli, Phys.\ Rev.\ D {\bf 42} 2638 (1990).
A.\ Connes, C.\ Rovelli, Class.\ Quant.\ Grav.\
{\bf 11}, 2899 (1994).
\bibitem{IS.94} C.J.\ Isham, J.\ Math.\ Phys.\  {\bf 35} (1994) 2157.
\bibitem{M.14} E.\ Moreva et al,
Phys.\ Rev.\ A {\bf 89}, 052122 (2014).
\bibitem{M.15}S.\ Massar et al,
Phys.\ Rev.\ A {\bf 92}, 030102 (2015).
\bibitem{FC.13}J.R. McClean, J.A.\ Parkhill, A.\ Aspuru-Guzik,
Proc.\ Natl.\ Ac.\ Sci.\ U.S.A.\ {\bf 110},  E3901 (2013); J.R. McClean,  A.\ Aspuru-Guzik,
Phys.\ Rev.\ A {\bf 91}, 012311 (2015).
\bibitem{CR.15}E.\ Castro-Ruiz, F.\ Giacomini, C,\ Brukner, arXiv:1507.01955 (2015).
\bibitem{Ve.14} V.\ Vedral, arXiv:1408.6965 (2014).
\bibitem{QT.15}V.\ Giovannetti, S.\ Lloyd, L.\ Maccone,  Phys.\ Rev.\ D {\bf 92}, 045033 (2015).
\bibitem{Ga.09} R. Gambini, R.A. Porto, J. Pullin, S. Torterolo,
Phys.\ Rev.\ D 79, 041501(R) (2009).7
\bibitem{NC.00} M.A.\ Nielsen, I.L.\ Chuang,
{\em Quantum Computation and Quantum Information}
(Cambridge Univ. Press, Cambridge, UK, 2000).
\bibitem{DW.67}B.S. DeWitt, Phys.\ Rev.\ {\bf 160}, 1113 (1967).
\bibitem{HH.83}J.B.\ Hartle, S.W.\ Hawking, Phys.\ Rev.\ D {\bf 28}, 2960 (1983).
\bibitem{S2.00} V.\ Coffman, J.\ Kundu,  W.K.\ Wootters, Phys.\ Rev.\ A {\bf 61}, 052306 (2000);
\bibitem{RC.03}P.\ Rungta, C.M.\ Caves, Phys.\ Rev.\ A {\bf 67}, 012307 (2003).
\bibitem{GR.14} N.\ Gigena, R.\ Rossignoli, Phys.\ Rev.\ A {\bf 90}, 042318 (2014).
\bibitem{T.13} T.\ Tanaka, G.\ Kimura, H.\ Nakazato, Phys.\ Rev.\ A {\bf 87}, 012303 (2013).
 H.\ Nakazato et al,  Phys.\ Rev.\ A {\bf 85} 042316 (2012).
\bibitem{Wo.97} S.\ Hill and  W.K.\ Wootters, Phys.\ Rev.\ Lett.\ {\bf 78}, 5022 (1997);
W.K.\ Wootters, Phys.\ Rev.\ Lett.\ {\bf 80}, 2245 (1998).
\bibitem{Bha.97} R.\ Bhatia, Matrix Analysis (Springer-Verlag, New York, 1997).
\bibitem{CR.02} N.\ Canosa, R.\ Rossignoli, Phys.\ Rev.\ Lett.\ 88, 170401 (2002);  
R.\ Rossignoli,  N.\ Canosa, Phys.\ Rev.\ A 67, 042302 (2003).
\bibitem{DCT.91} A.\ Dembo, T.M.\ Cover, J.A.\ Thomas,  IEEE
Trans.\ on Inf.\ Th.\ {\bf 37}, 1501  (1991).
\bibitem{PDO.01} T.\ Przebinda, V.\ DeBrunner,  M. \"Ozaydin,
IEEE Trans.\ on Inf.\ Th.\ {\bf 47}, 2086  (2001);
M.\ \"Ozaydin, T.\ Przebinda, J.\ Func.\ An.\ {\bf 215}, 241 (2004).
\bibitem{Hi.57} L.I.\ Hirschman, Amer. J. Math. {\bf 79}, 152 (1957).
\bibitem{DS.89} D.L.\ Donoho, P.B.\ Stark, SIAM J.\ Appl.\ Math.\ {\bf 49}, 906 (1989).
\bibitem{MS.08} S.\ Massar, P.\ Spindel, Phys.\ Rev.\ Lett.\ {\bf 100}, 190401 (2008).
\bibitem{RF.89} R.F.\ Werner, Phys.\ Rev.\ A {\bf 40}, 4277 (1989).
\bibitem{BV.96} C.H. Bennett, D.P. DiVincenzo, J.A. Smolin, W.K. Wootters, Phys.\ Rev.\ A
{\bf 54}, 3824 (1996).
\bibitem{DZ.01} H.\ Ollivier, W.H.\ Zurek, Phys.\ Rev.\ Lett.\ {\bf 88}, 017901 (2001);
\bibitem{HV.03} L.\ Henderson and V.\ Vedral, J.\ Phys.\ A {\bf 34}, 6899 (2001);
 V. Vedral, Phys.\ Rev.\ Lett.\ {\bf 90}, 050401 (2003).
\bibitem{KM.12}K.\ Modi et al, Rev.\ Mod.\ Phys.\ {\bf 84}, 1655 (2012).
\bibitem{RCC.10}R.\ Rossignoli, N.\ Canosa, L.\ Ciliberti, Phys.\ Rev.\ A {\bf 82}, 052342 (2010).
\bibitem{DP.04} A.\ C.\ Doherty, P.A.\ Parrilo, F.M.\ Spedalieri, Phys.\ Rev.\ A {\bf 69}, 022308 (2004).
\bibitem{YH.14} Y. Huang, New.\ J.\ Phys.\ {\bf 16}, 033027 (2014).
\bibitem{AA.90}J. Anandan and Y. Aharonov, Phys. Rev. Lett.\ {\bf 65}, 1697 (1990).
\bibitem{OV.06}T.J.\ Osborne, F.\ Verstraete, Phys.\ Rev.\ Lett.\ {\bf 96}, 220503 (2006).
\bibitem{Wi.12} F.\ Wilczek, Phys.\ Rev.\ Lett.\ {\bf 109}, 160401 (2012).
\end{thebibliography}
\end{document}